\documentclass[aps,prl,showkeys,superscriptaddress,singlecolumn,nofootinbib,floatfix]{revtex4-2}

\usepackage{subfiles}
\usepackage{amsmath}
\usepackage{amssymb}
\usepackage{amsthm}
\usepackage{mathrsfs}
\usepackage{graphicx}
\usepackage{epstopdf}
\usepackage{fancyhdr}
\usepackage{array}
\usepackage[all]{xy}
\usepackage{eufrak}
\usepackage{euscript}
\usepackage{enumerate}
\usepackage{slashed}
\usepackage{hyperref}
\usepackage{subfigure} % NEEDED IN ORDER TO USE SUBFIG
\usepackage{epstopdf} % COMPILE EPS (AND PDF) FIGURES USING PDFLATEX!!!
\hypersetup{pdftex,colorlinks=true,linkcolor=blue,citecolor=blue,menucolor=black,urlcolor=blue,filecolor=blue}

\newcommand{\mysection}[1]{{\vspace{10 pt}\noindent \emph{{\textbf{#1}}--}}}

\begin{document}

\title{Violation of energy conditions and entropy production in holographic Bjorken flow}

\author{Romulo Rougemont}
\email{romulo.pereira@uerj.br}
\affiliation{Departamento de F\'{i}sica Te\'{o}rica, Universidade do Estado do Rio de Janeiro,
Rua S\~{a}o Francisco Xavier 524, 20550-013, Maracan\~{a}, Rio de Janeiro, Rio de Janeiro, Brazil}

\author{Jorge Noronha}
\email{jn0508@illinois.edu}
\affiliation{Illinois Center for Advanced Studies of the Universe\\ Department of Physics, 
University of Illinois at Urbana-Champaign, Urbana, IL 61801, USA}

\author{Willians Barreto}
\email{willians.barreto@ufabc.edu.br}
\affiliation{Departamento de F\'{i}sica Te\'{o}rica, Universidade do Estado do Rio de Janeiro, Rua S\~{a}o Francisco Xavier 524, 20550-013, Maracan\~{a}, Rio de Janeiro, Rio de Janeiro, Brazil}
\affiliation{Centro de Ci\^{e}ncias Naturais e Humanas, Universidade Federal do ABC, Av. dos Estados 5001, 09210-580 Santo Andr\'{e}, S\~{a}o Paulo, Brazil}
\affiliation{Centro de F\'{i}sica Fundamental, Universidad de Los Andes, M\'{e}rida 5101, Venezuela}

\author{Gabriel S.~Denicol}
\email{gsdenicol@id.uff.br}
\affiliation{Instituto de F\'isica, Universidade Federal Fluminense, UFF, Niter\'oi, 24210-346, RJ, Brazil}

\author{Travis Dore}
\email{tdore2@illinois.edu}
\affiliation{Illinois Center for Advanced Studies of the Universe\\ Department of Physics, 
University of Illinois at Urbana-Champaign, Urbana, IL 61801, USA}

\begin{abstract}
We demonstrate that a Bjorken expanding strongly coupled $\mathcal{N}=4$ Supersymmetric Yang-Mills plasma can display dynamically-driven violations of the dominant and also the weak energy condition during hydrodynamization. In addition, we find that a period of vanishing entropy production in far-from-equilibrium stages induces later violations of the dominant energy condition in the strongly coupled plasma. Such violations cannot occur in a classical description of hydrodynamization and suggest that the inclusion of quantum effects in transport can lead to new phenomena  in these regimes, even for systems without anomalies or spin.

%This result suggests that the inclusion of quantum effects in rapidly expanding systems can lead to new phenomena in the far-from-equilibrium regime.

%We unveil on a new critical feature of the hydrodynamization process in $\mathcal{N}=4$ Supersymmetric Yang-Mills plasma undergoing Bjorken flow -- the direct connection between a vanishing entropy production and the violation of energy conditions far from equilibrium.

%We find that a period of vanishing entropy production in far-from-equilibrium stages always acts as a precursor of later violations of the dominant energy condition in the strongly coupled plasma.

%We also establish for the first time a direct connection between entropy production and the violation of energy conditions in relativistic many-body quantum systems far from equilibrium. We find that a period of vanishing entropy production in far-from-equilibrium stages always acts as a precursor of later violations of the dominant energy condition in the strongly coupled plasma.

%Our paper points out for the first time a new critical feature of the hydrodynamization process, i.e., the direct connection between a vanishing entropy production and the violation of energy conditions far from equilibrium. 
\end{abstract}

\maketitle

%%%%%%%%%%%%%%%%%%%%%%%%%%%%%%%%%

\mysection{Introduction} The quark-gluon plasma (QGP) formed in ultrarelativistic heavy-ion collisions \cite{Heinz:2013th} provides a way to investigate the out of equilibrium properties of quantum chromodynamics \cite{Berges:2020fwq}. Driven by the observation of hydrodynamic signatures in small-system proton-proton and proton-nucleus collisions \cite{Chatrchyan:2013nka,Abelev:2014mda,Aaboud:2017acw,PHENIX:2018lia}, the focus has recently evolved towards understanding how hydrodynamic behavior may emerge even in the far-from-equilibrium regime. The current working hypothesis is that hydrodynamics may be defined as a universal attractor \cite{Heller:2015dha} in which dissipative contributions can still display universal behavior even when local gradients are large \cite{Florkowski:2017olj}. Holography \cite{Maldacena:1997re,Gubser:1998bc,Witten:1998qj,Witten:1998zw} provides a very natural framework to investigate this problem, as it allows for comprehensive real-time calculations of the onset of hydrodynamics in strongly coupled gauge theories starting from states arbitrarily far from equilibrium, see e.g. \cite{Chesler:2008hg,Fuini:2015hba,Critelli:2017euk,Cartwright:2019opv,Chesler:2009cy,Heller:2011ju,Heller:2012je,Chesler:2013lia,vanderSchee:2014qwa,Jankowski:2014lna,Romatschke:2017vte,Spalinski:2017mel,Florkowski:2017olj,Casalderrey-Solana:2017zyh,Critelli:2018osu,Kurkela:2019set,Chesler:2010bi,Casalderrey-Solana:2013aba,Chesler:2015bba,Grozdanov:2016zjj,Attems:2016tby,Casalderrey-Solana:2016xfq,Attems:2017zam,Attems:2018gou}.

Currently, in heavy-ion collision simulations the hydrodynamization of the hot and dense QGP is modeled using effective kinetic descriptions \cite{Kurkela:2014tea,Kurkela:2018wud,Almaalol:2020rnu}, which describe the QGP as a dilute gas of classical weakly-coupled quasiparticles \cite{Arnold:2002zm}. This is in sharp contrast \cite{Ghiglieri:2018dgf} to holography, where hydrodynamization takes place without relying on the quasiparticle concept. These two descriptions also display qualitatively different behavior in their approach to hydrodynamics. Kinetic descriptions show purely exponential decay of nonhydrodynamic modes \cite{Denicol:2011fa}, while in holography these modes also feature an oscillatory behavior \cite{Kovtun:2005ev,Denicol:2011fa,Heller:2014wfa,Florkowski:2017olj}.

Another fundamental difference is that kinetic models based on the Boltzmann equation always satisfy a set of energy conditions \cite{HawkingEllisBook,WaldBookGR1984} because the single particle distribution function is non-negative \cite{degroot}, while violations can be found in some holographic settings \cite{Figueras:2012rb,Arnold:2014jva}. For instance, the weak energy condition (WEC) guarantees the existence of a non-negative energy density for all observers by imposing that the energy-momentum tensor, $\langle T_{\mu\nu}\rangle$, satisfies $\langle T_{\mu\nu}\rangle t^\mu t^\nu\ge 0$ for \emph{all} timelike vectors $t^\mu$. Also, the dominant energy condition (DEC) guarantees that the speed of energy flow of matter is subluminal, by imposing that for \emph{all} future directed timelike vectors $t^\mu$ the vector $ -\langle T^{\mu}_\nu\rangle t^\nu$ must also be a future directed timelike or null vector \cite{WaldBookGR1984}. These conditions are expected to be satisfied by physically reasonable classical matter \cite{WaldBookGR1984} and, thus, their violations would indicate the presence of nontrivial quantum phenomena. In transport approaches, such violations can only possibly appear outside the classical regime due to quantum effects \cite{Baym:1961zz}.  

In this work we show that these energy conditions can be violated in the far-from-equilibrium strongly coupled $\mathcal{N}=4$ Supersymmetric Yang-Mills (SYM) plasma undergoing Bjorken flow \cite{Bjorken:1982qr}, which is a widely used toy model of heavy-ion collisions. Such violations may occur throughout the onset of hydrodynamics, even when the initial conditions satisfy these constraints.  We go beyond previous works and also investigate the non-equilibrium entropy density, defined by the area of the apparent horizon, and provide for the first time clear numerical evidence that a transiently vanishing far-from-equilibrium entropy production occurs and its presence precedes an imminent or effective violation of the DEC. These results imply that in holography the hydrodynamization process contains quantum effects which, as such, are beyond the reach of any classical approach. Therefore, comparisons between holography and kinetic theory approaches can only be meaningfully done after energy condition violations have ceased.

%%%%%%%%%%%%%%%%%%%%%%%%%%%%%%%%%
\mysection{Energy conditions in Bjorken flow} A toy model commonly used to describe the expansion of the matter formed in ultrarelativistic heavy-ion collisions near mid-rapidity \cite{Arsene:2004fa,Adcox:2004mh,Back:2004je,Adams:2005dq,Aad:2013xma} is Bjorken flow \cite{Bjorken:1982qr}. The assumptions behind it are boost invariance along the beamline, plus translation and rotation invariance in the transverse plane (i.e., no motion in the $xy$ plane). The longitudinally expanding system moves at the speed of light in the $z$ direction and the motion is more naturally described in Milne coordinates $(\tau,\xi,x,y)$, where $\tau=\sqrt{t^2-z^2}$ and $\xi=\frac{1}{2} \ln\left[ (t+z)/(t-z)\right]$ are the propertime and the spacetime rapidity, respectively (we use natural units $\hbar = c = k_B = 1$). In terms of these coordinates, the 4d Minkowski metric becomes $ds^2_{\textrm{(4d)}} = -d\tau^2+\tau^2 d\xi^2+dx^2+dy^2$ and the local 4-velocity of the expanding system is $u^\mu = (1,0,0,0)$ (assuming invariance under reflections $\xi \to -\xi$), which implies that the system's expansion rate is $\nabla_\mu u^\mu = 1/\tau$, and the shear tensor is given by $\sigma_{\mu\nu} =\mathrm{diag}(0,4/3\tau,-2/3\tau,-2/3\tau)$. In a Bjorken expanding system, the energy-momentum tensor $\langle T^{\mu}_\nu\rangle = \mathrm{diag}(\varepsilon,p_L,p_T,p_T)$, where $\varepsilon$ is the energy density, while $p_L=\varepsilon/3+\pi_\xi^\xi$ and $p_T=\varepsilon/3-\pi_\xi^\xi/2$ are the longitudinal and transverse pressures, respectively, with $\pi^\mu_\nu$ being the shear-stress tensor. 

In Bjorken flow, the energy conditions lead to a set of constraints on the spacetime evolution of the system. For a conformal field theory (as e.g. the SYM plasma), the weak energy condition implies\footnote{For a conformal fluid  the strong energy condition, $\langle T_{\mu\nu}\rangle t^\mu t^\nu\ge -\langle T_\mu^\mu\rangle /2$ \cite{WaldBookGR1984}, is trivially equivalent to WEC.} \cite{Janik:2005zt} $\varepsilon(\tau)\ge 0$, $\partial_\tau\varepsilon(\tau)\le 0$, and $\tau\partial_\tau\ln\varepsilon(\tau)\ge -4$. In terms of the pressure anisotropy, $\Delta p \equiv p_T -p_L$, the two last inequalities give $-4\le \Delta p/ \varepsilon\le 2$. On the other hand, the dominant energy condition implies that $\varepsilon>0$ and $-1 \leq \Delta p/\varepsilon \leq 2$. In the next sections we explain how the energy conditions can be analyzed in holography.

%%%%%%%%%%%%%%%%%%%%%%%%%%%%%%%%%
\mysection{Gravitational description}
The first full numerical approach to the holographic Bjorken flow of the SYM plasma was originally formulated in  \cite{Chesler:2009cy}, with further developments presented in Refs.\ \cite{Heller:2011ju,Heller:2012je,Chesler:2013lia,vanderSchee:2014qwa,Jankowski:2014lna}. The dual description only involves 5d classical gravity in asymptotically AdS$_5$ spacetime. We follow \cite{Chesler:2013lia} and use a characteristic formulation of general relativity closely related to \cite{Bondi:1962px,Sachs:1962wk,Winicour:1998tz,Lehner:2001wq,Winicour:2013gha}, where the Ansatz for the 5d bulk metric field compatible with diffeomorphism invariance and the Bjorken flow symmetries can be written in the infalling Eddington-Finkelstein (EF) coordinates as follows \cite{Chesler:2009cy,Chesler:2013lia}
\begin{equation}
ds^2 =2d\tau\left[dr-A(\tau,r) d\tau \right]+\Sigma(\tau,r)^2\left[e^{-2B(\tau,r)}d\xi^2 + e^{B(\tau,r)}(dx^2+dy^2)\right],
\label{lineElement}
\end{equation}
where $r$ is the radial holographic direction with boundary at $r\to\infty$. Einstein's equations give the following set of coupled $1+1$ PDE's for the metric coefficients $A(r,\tau)$, $B(r,\tau)$, and $\Sigma(r,\tau)$
\begin{subequations}
\begin{align}
\Sigma''+\frac{\Sigma B'\,^2}{2} &=0, \label{pde1}\\
(d_+\Sigma)'+\frac{2\Sigma ' d_+\Sigma}{\Sigma} - 2\Sigma &=0, \label{pde2}\\
\Sigma(d_+B)' +\frac{3(B' d_+\Sigma+\Sigma 'd_+B)}{2} &=0, \label{pde3}\\
A''+\frac{4+3B' d_+B - 12(\Sigma ' d_+\Sigma)/\Sigma^2}{2} &=0, \label{pde4}\\
d_+(d_+\Sigma)+\frac{\Sigma (d_+B)^2}{2} -A' d_+\Sigma &=0, \label{pde5}
\end{align}
\end{subequations}
where $'\equiv\partial_r$ is the directional derivative along infalling radial null geodesics and $d_{+}\equiv \partial_{\tau}+A(r,\tau)\partial_r$ is the directional derivative along outgoing radial null geodesics. Equation \eqref{pde5} is a constraint that can be used in order to check the precision of the numerical solutions obtained by solving the nested equations \eqref{pde1} --- \eqref{pde4}. The line element \eqref{lineElement} has residual diffeomorphism invariance under radial shifts $r\mapsto r+\lambda(\tau)$, with $\lambda(\tau)$ being an arbitrary function of time. When numerically solving Einstein's equations, we adopt the procedure discussed in \cite{Chesler:2013lia} and fix different values for $\lambda(\tau)$ on different time slices by requiring that the radial position of the apparent horizon remains fixed.

By imposing that the 4d line element $ds^2_{\textrm{(4d)}}$ is recovered from the 5d metric \eqref{lineElement} at the boundary (up to the conformal factor $r^2$ of AdS$_5$), the UV near-boundary expansions of the bulk metric coefficients assume the following form \cite{Chesler:2009cy,Critelli:2018osu}
\begin{equation}
A(r,\tau) = \frac{(r+\lambda(\tau))^2}{2}-\partial_\tau\lambda(\tau)+\sum_{n=1}^\infty \frac{a_n(\tau)}{r^n}, \qquad B(r,\tau) = -\frac{2\ln(\tau)}{3}+\sum_{n=1}^\infty \frac{b_n(\tau)}{r^n}, \qquad \Sigma(r,\tau) = \tau^{1/3}r+\sum_{n=0}^\infty \frac{s_n(\tau)}{r^n}.
\end{equation}
By substituting the above UV expansions back into Einstein's equations, and solving the resulting algebraic equations order by order in $r$, one can find the UV coefficients $\{a_n(\tau),b_n(\tau),s_n(\tau)\}$ as functions of $\tau$, $\lambda(\tau)$, $a_2(\tau)$, and its derivatives. The initial data needed to solve Einstein's equations are defined by the value of $a_2(\tau)$ at the initial time $\tau_0$ and the initial profile for the metric anisotropy $B(r,\tau_0)$ ($\lambda(\tau_0)$ must also be specified). The nonzero components of the energy-momentum tensor are given by the (normalized) energy density $\hat{\varepsilon}(\tau) \equiv \kappa_{5}^{2}\langle T_{\tau\tau} \rangle = -3a_2(\tau)$, $\hat{p}_T(\tau) \equiv \kappa_{5}^{2}\langle T_x^x \rangle = -3a_2(\tau)-\frac{3}{2}\tau \partial_\tau a_2(\tau)$, and $\hat{p}_L(\tau) \equiv \kappa_{5}^{2}\langle T_\xi^\xi \rangle =  3a_2(\tau)+3\tau \partial_\tau a_2(\tau)$, where $\kappa_5^2= 4\pi^2/N_c^2$  and $N_c$ is the number of colors. One needs to determine the time evolution of the dynamical UV coefficient $a_2(\tau)$ in order to obtain the time evolution of these physical observables. This can be most easily done by defining a new holographic coordinate $u=1/r$ and also subtracted bulk fields $X\mapsto X_s$, $X\in\{B,\Sigma,d_+\Sigma,d_+B,A\}$, from which one may simply extract $a_2(\tau) = A_s(u=0,\tau)$ (see the Supplemental Material for details).

In the presence of a black hole apparent horizon, the portion of the bulk geometry within the horizon is causally disconnected from observers at the boundary.
For the line element in \eqref{lineElement}, the radial position of the apparent horizon, $r_{\textrm{AH}}(\tau)$, may be determined \cite{Chesler:2013lia} by finding the value of the radial coordinate which solves (see the Supplemental Material for details)
\begin{align}
d_+\Sigma(r_{\textrm{AH}}(\tau),\tau)=0.
\label{AppHor}
\end{align}

Finally, concerning the initial conditions in the bulk, we set \cite{Critelli:2018osu} $\lambda(\tau_0) = 0$,
\begin{equation}
B_s(u,\tau_0) = \Omega_1 \cos(\gamma_1 u) + \Omega_2 \tan(\gamma_2 u) + \Omega_3 \sin(\gamma_3 u) + \sum_{i=0}^{5}\beta_i u^i  +\, \frac{\alpha}{u^4} \left[-\frac{2}{3} \ln\left(1+ \frac{u}{\tau _0}\right) + \frac{2 u^3}{9 \tau_0^3} - \frac{u^2}{3 \tau _0^2}+\frac{2 u}{3 \tau _0}\right],\label{Bs0}
\end{equation}
and choose values for the parameters $\{\Omega_i,\gamma_i,\beta_i,\alpha\}$, and for $a_2(\tau_0)$ (which sets the initial energy density of the fluid), as given in Table \ref{tabICs} of the Supplemental Material. We take $\tau_0=0.2$ as the initial time of our numerical simulations.

%%%%%%%%%%%%%%%%%%%%%%%%%%%%%%%%%
\mysection{Energy condition violations} In this section we present our results for the pressure anisotropy. As a consistency check of our numerical solutions, we compare them with known hydrodynamical limits obtained up to second-order in the gradient expansion \cite{Baier:2007ix,Romatschke:2017vte}, $\left[\Delta \hat{p}/{\hat{\varepsilon}}\right]_{\textrm{2nd}} = {2}/({3\pi\omega_\Lambda}) + {2\,(1-\ln 2)}/({9\pi^2\omega_\Lambda^2})$, where $\omega_\Lambda(\tau)\equiv \tau\, T_{\textrm{eff}}(\tau)$ is an effective dimensionless time measure, with $T_{\textrm{eff}}(\tau)$ being an effective temperature, which is defined in terms of an energy scale $\Lambda$ using a late time expansion of the energy density \cite{Chesler:2009cy,Florkowski:2017olj} (see the Supplemental Material for details).

\begin{figure*}%[h]
\center
\subfigure[]{\includegraphics[width=0.47\textwidth]{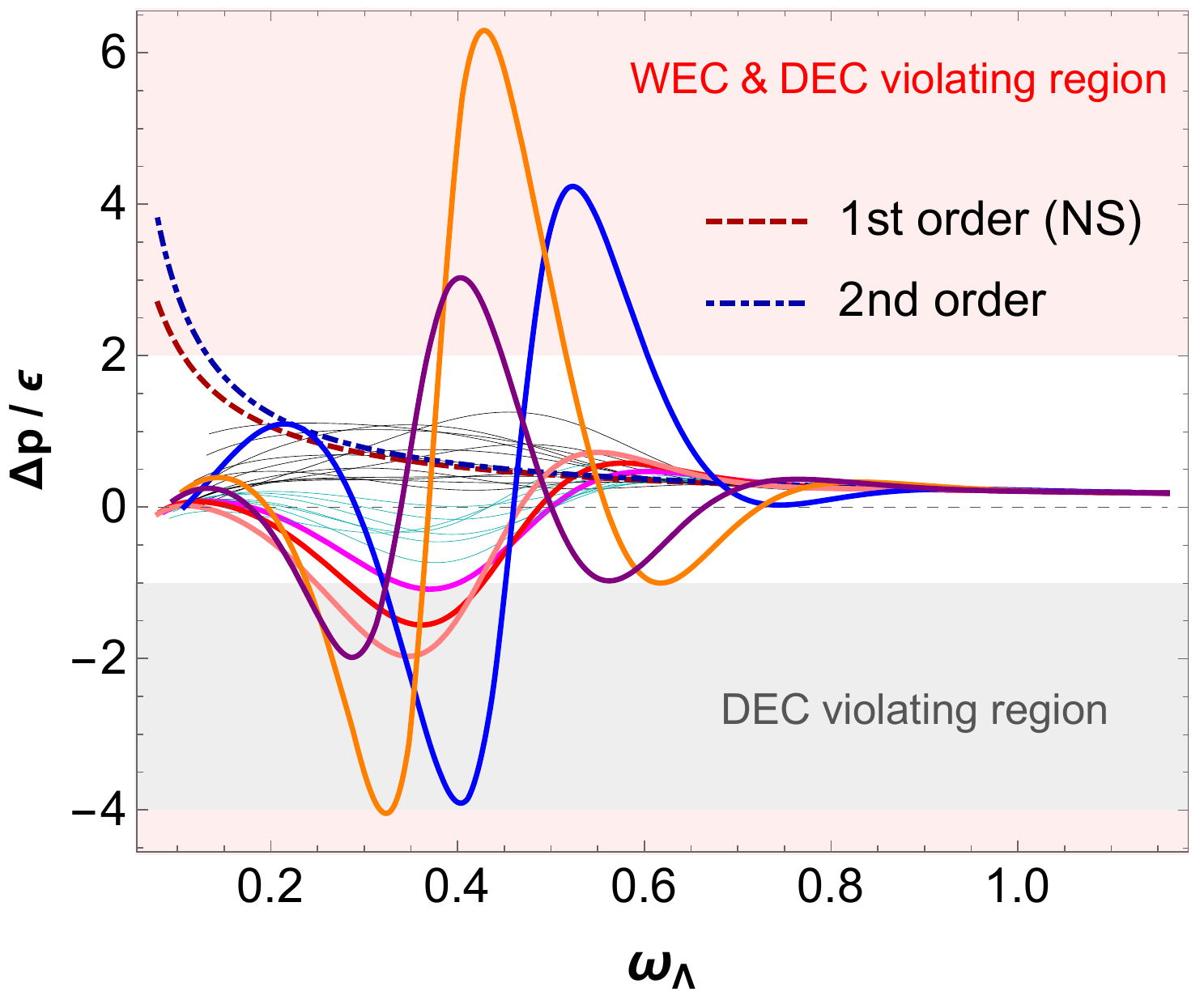}}
\qquad
\subfigure[]{\includegraphics[width=0.47\textwidth]{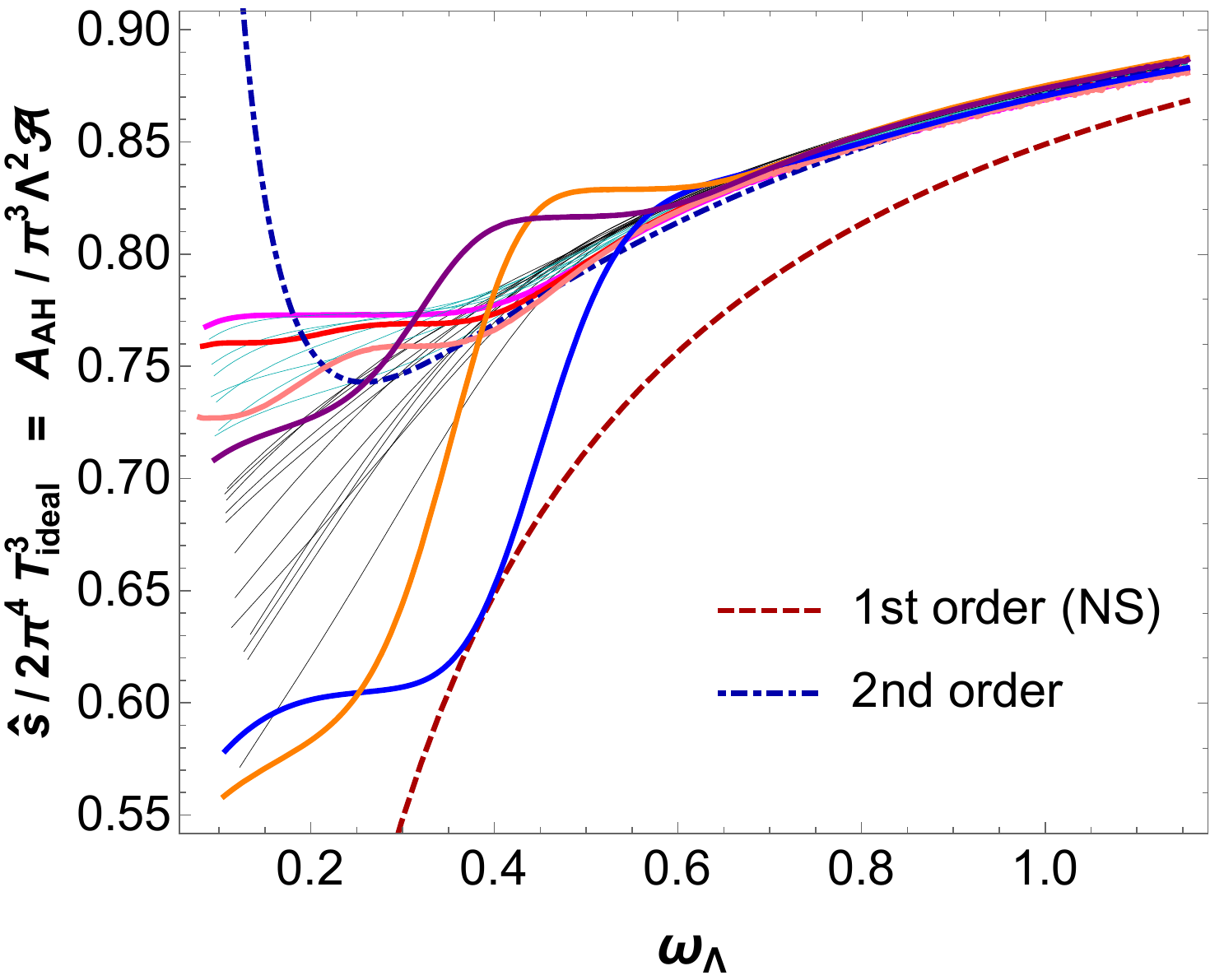}}
\caption{(a) Pressure anisotropy, (b) non-equilibrium entropy density, and their late time hydrodynamic approximations. Solid colored lines are the solutions that violate energy conditions.}
\label{fig:result1}
\end{figure*}

A comparison between the full numerical evolution of the pressure anisotropy for several initial conditions and the corresponding hydrodynamic results at first and second order in the gradient expansion is shown in Fig.\ \ref{fig:result1} (a). We see that the pressure anisotropy of the holographic numerical solutions, although highly dependent on the chosen initial condition at early times, indeed converges at late times to the corresponding hydrodynamic results for all the initial conditions. Our results agree with previous studies in the literature \cite{Chesler:2009cy,Heller:2011ju,Heller:2012je,Jankowski:2014lna,Romatschke:2017vte,Spalinski:2017mel,Florkowski:2017olj,Casalderrey-Solana:2017zyh,Critelli:2018osu,Kurkela:2019set}, which also found that this effective hydrodynamic description holds even in the presence of a sizable pressure anisotropy.

We show in Fig.\ \ref{fig:result1} (a) that there are solutions that satisfy both DEC and also WEC at the initial time but later develop violations of such conditions in the far-from-equilibrium regime before the system hydrodynamizes (see the full colored thick curves). Therefore, at such earlier times, the system is inherently non-hydrodynamical and any attempt to match its behavior to hydrodynamics is misleading. It was previously known e.g. that WEC can be violated in holographic shock wave collisions \cite{Arnold:2014jva}.  Our results show that DEC and WEC can be violated even in the simplest holographic toy model of heavy-ion collisions (i.e. Bjorken flow) and, thus, such violations are not rare in strongly-coupled quantum fluids. Therefore, there is the possibility that they are important ingredients to understand the emergence of hydrodynamic behavior in the QGP formed in heavy-ion collisions.  

%%%%%%%%%%%%%%%%%%%%%%%%%%%%%%%%%
\mysection{Holographic entropy out of equilibrium} The Bekenstein-Hawking relation \cite{Bekenstein:1973ur,Hawking:1974sw} associates the thermodynamical entropy of a black hole in equilibrium with the area of its event horizon. However, in out-of-equilibrium settings, it was argued in \cite{Figueras:2009iu} that the holographic non-equilibrium entropy should be associated with the area of the apparent horizon instead of the area of the event horizon. Indeed, the holographic non-equilibrium entropy has been considered in many other works \cite{Chesler:2009cy,Heller:2011ju,Heller:2012je,Jankowski:2014lna,vanderSchee:2014qwa,Grozdanov:2016zjj,Buchel:2016cbj,Muller:2020ziz,Engelhardt:2017aux} as being related to the area of the apparent horizon. The apparent horizon lies behind the event horizon and converges to the latter at late times and, thus, for sufficiently long times the areas of both horizons coincide giving the same result for the entropy in equilibrium. See also \cite{Jansen:2016zai,Jansen:2020ign} for further discussions on the properties of the non-equilibrium entropy and the behavior of apparent and event horizons.

The area of the apparent horizon in  holographic Bjorken flow is given by $A_{\textrm{AH}}(\tau) = |\Sigma(u_{\textrm{AH}},\tau)|^3 \mathcal{A}$, where $\mathcal{A} = \int dx dy d\xi$ and $\tau \mathcal{A}$ is the spatial volume in Milne coordinates. Using the Bekenstein-Hawking relation for the apparent horizon area to define the non-equilibrium holographic entropy, one finds that the (normalized) entropy density is given by $\hat{s}(\tau)\equiv\kappa_5^2 S(\tau)/\tau\mathcal{A}= 2\pi A_{\textrm{AH}}(\tau)/\tau\mathcal{A} =2\pi|\Sigma(u_{\textrm{AH}},\tau)|^3/\tau$. The late time expansions for the area of the apparent horizon of the gravity dual of a Bjorken expanding SYM plasma can be found analytically. Following \cite{Kinoshita:2008dq,Nakamura:2006ih,Figueras:2009iu,Chesler:2009cy}, the holographic entropy density to second-order in the gradient expansion, now normalized by the asymptotic ideal effective temperature $T_{\textrm{ideal}}(\tau)= \Lambda^{2/3}/\tau^{1/3}$, is given by $\hat{s}_{\textrm{2nd}}/2\pi^4 T^3_{\textrm{ideal}}=1-1/2\pi(\Lambda\tau)^{2/3}+(2+\pi+\ln 2)/24\pi^2 (\Lambda\tau)^{4/3}$.

In Fig.\ \ref{fig:result1} (b) we plot our numerical results for the dimensionless quantity $\hat{s}/2\pi^4 T^3_{\textrm{ideal}}=|\Sigma(u_{\textrm{AH}},\tau)|^3/\pi^3\Lambda^2$, which is normalized to unity in equilibrium. Differently than in Fig.\ \ref{fig:result1} (a), the first-order Navier-Stokes (NS) and second-order hydrodynamic curves only converge for values of $\omega_\Lambda$ considerably larger than unity. Also, one can see that the second-order hydrodynamic curve becomes a good approximation to the numerical solutions much earlier than the NS result, while for the pressure anisotropy this is not the case. 

More importantly, we note the presence of plateaus for certain solutions displayed in Fig.\ \ref{fig:result1} (b), which describe stages with zero entropy production far from equilibrium. There is a physical connection between the plateaus in Fig.\ \ref{fig:result1} (b) and the pressure anisotropy in Fig.\ \ref{fig:result1} (a). When a single plateau is formed (see e.g. the orange curve) indicating zero entropy production, the normalized entropy in Fig.\ \ref{fig:result1} (b) only starts to increase close to the hydrodynamization time associated with 2nd order hydrodynamics. Similarly, if more plateaus are present (see e.g. the salmon curve), the normalized entropy only starts to increase again after the last plateau gets close to the curve corresponding to 2nd order hydrodynamics. We also observe that in such cases the approach to 2nd order hydrodynamics occurs if and only if a local minimum, with $\Delta p/\epsilon \le -1$, is observed for the pressure anisotropy \emph{after} the first time  such a plateau has been produced. Moreover, in these cases, the formation of a single plateau for $\hat{s}/2\pi^4T^3_{\textrm{ideal}}$ \emph{later} implies a local minimum with $\Delta p/\epsilon = -1$ (corresponding to the boundary to DEC violation), while the formation of multiple plateaus \emph{later} implies a local minimum with $\Delta p/\epsilon < -1$ (corresponding to effective DEC violation). On the other hand, we point out that there are solutions which violate DEC and WEC without displaying plateaus for $\hat{s}/2\pi^4T^3_{\textrm{ideal}}$ far from equilibrium (see e.g. the blue curve). Particularly, we have observed no peculiar features regarding $\hat{s}/2\pi^4T^3_{\textrm{ideal}}$ associated with violations of DEC and WEC for values of $\Delta p/\epsilon > 2$. 

The findings described above provide clear numerical evidence, for the first time, that states with transiently vanishing far-from-equilibrium entropy production occur in $\mathcal{N}=4$ SYM undergoing Bjorken flow and their presence precede an imminent or effective violation of the dominant energy condition. 

\mysection{Conclusions} In this paper we showed that the evolution of the strongly coupled SYM plasma undergoing Bjorken flow can  transiently violate the dominant and weak energy conditions even when the initial data satisfy these constraints. We also established for the first time a direct connection between entropy production and the violation of energy conditions in relativistic many-body quantum systems, by providing clear numerical evidence that a transiently vanishing entropy production in far from equilibrium stages is always a precursor of later imminent or effective DEC violation in the Bjorken expanding SYM plasma. This is in sharp contrast with free streaming in classical kinetic models, where the entropy production is zero but the energy conditions are always satisfied.

Our findings unveil new aspects of the hydrodynamization process of far from equilibrium quantum media and may have also important consequences to the understanding of the later emergence of hydrodynamic behavior in heavy-ion collisions. Assuming $\Lambda \sim 0.2$ GeV, the region where energy condition violations occur and the out-of-equilibrium entropy may display transient plateaus ($\omega_\Lambda \lesssim 0.6$) is mapped into times $\tau \lesssim 0.5$ fm/c, which is within the so-called pre-hydrodynamic regime of heavy-ion collisions \cite{Kurkela:2018wud} where the QCD matter is expected to transition from a classical field dominated, gluon-saturated regime \cite{McLerran:1993ni,Gale:2012rq} to a hydrodynamic medium. Therefore, if the energy conditions may be indeed violated in the early stages of heavy-ion collisions, the description of the pre-hydrodynamic phase will be inherently outside the regime of validity of kinetic theory approaches. Such a possibility is also a motivation for investigating quantum effects in the context of the Kadanoff-Baym equations \cite{Baym:1961zz}, which may possibly lead to new insights into the hydrodynamization process in heavy-ion collisions.

It would be interesting to check if the features found here are also displayed by other holographic models when considered in the far-from-equilibrium regime, especially realistic holographic constructions designed for phenomenological applications, such as e.g. \cite{Finazzo:2016mhm,Critelli:2017oub,Grefa:2021qvt}. One may also investigate how these features manifest in the behavior of nonlocal operators \cite{Pedraza:2014moa,DiNunno:2017obv}, and in the presence of external electromagnetic fields \cite{Fuini:2015hba,Cartwright:2019opv,Cartwright:2020qov,Ghosh:2021naw}, which are expected to be relevant in the early stages of high energy peripheral heavy-ion collisions.

%%%%%%%%%%%%%%%%%%%%%%%%%%%%%%%%%
\mysection{Acknowledgments} The authors thank R.~Critelli for the collaboration during the first stages of this work and M.~Heller and M.~Spalinski for very insightful comments on an early version of this paper. R.R. and W.B. acknowledge financial support by Universidade do Estado do Rio de Janeiro (UERJ) and Funda\c{c}\~{a}o Carlos Chagas Filho de Amparo \`{a} Pesquisa do Estado do Rio de Janeiro (FAPERJ). J.N. is partially supported by the U.S. Department of Energy, Office of Science, Office for Nuclear Physics under Award No.\ DE-SC0021301. G.S.D. is partially supported by FAPERJ, grant number E-26/202.747/2018, and Conselho Nacional de Desenvolvimento Cient\'ifico e Tecnol\'ogico (CNPq). J.N. and G.S.D. also thank Funda\c{c}\~{a}o de Amparo \`{a}  Pesquisa do Estado de S\~ao Paulo (FAPESP), grant number 2017/05685-2. T.D. is supported by the US-DOE Nuclear Science Grant No. DESC0020633.

%%%%%%%%%%%%%%%%%%%%%%%%%%%%%%%%%%
\appendix
\section*{Supplemental Material}
\label{sec:HolNum}

In this Supplemental Material we give further details about the numerical procedure used to obtain the results in the main text.

Concerning the definition of the dimensionless time measure $\omega_\Lambda(\tau) = \tau T_{\textrm{eff}}(\tau)$, we follow \cite{Florkowski:2017olj} and take $T_{\textrm{eff}}(\tau)$ to be given by the third-order hydrodynamic truncation for the energy density of the SYM plasma \cite{Booth:2009ct},
\begin{equation}
T_{\textrm{3rd}}(\tau) = \frac{\Lambda}{(\Lambda\tau)^{1/3}} \left[ 1 - \frac{1}{6\pi(\Lambda\tau)^{2/3}} + \frac{-1+\ln 2}{36\pi^2(\Lambda\tau)^{4/3}} +\, \frac{-21 + 2\pi^2 + 51\ln 2 - 24\ln^2 2}{1944\pi^3 (\Lambda\tau)^2} \right].
\label{eq:time}
\end{equation}
In our numerical simulations, the energy scale $\Lambda$ is extracted for each initial condition through a fit of the late time result for the full numerical energy density to its corresponding NS result \cite{Janik:2005zt,Chesler:2009cy,Critelli:2018osu}.

The set of initial conditions analyzed in the present work is provided in table \ref{tabICs}.
\begin{table}[h]
\centering
\begin{tabular}{|c||c|c|c|c|c|c|c|c|c|c|c|c|c||c|}
\hline
IC$\#$ & $\Omega_1$ & $\gamma_1$ & $\Omega_2$ & $\gamma_2$ & $\Omega_3$ & $\gamma_3$ & $\beta_0$ & $\beta_1$ & $\beta_2$ & $\beta_3$ & $\beta_4$ & $\beta_5$ & $\alpha$ & $a_2(\tau_0)$ \\
\hline
\hline
1 & 0 & 0 & 0 & 0 & 0 & 0 & 0.5 & -0.5 & 0.4 & 0.2 & -0.3 & 0.1 & 1 & -20/3 \\
\hline
2 & 0 & 0 & 0 & 0 & 0 & 0 & 0.2 & 0.1 & -0.1 & 0.1 & 0.2 & 0.5 & 1.02 & -20/3 \\
\hline
3 & 0 & 0 & 0 & 0 & 0 & 0 & 0.1 & -0.5 & 0.5 & 0 & 0 & 0 & 1 & -20/3 \\
\hline
4 & 0 & 0 & 0 & 0 & 0 & 0 & 0.1 & 0.2 & -0.5 & 0 & 0 & 0 & 1 & -20/3 \\
\hline
5 & 0 & 0 & 0 & 0 & 0 & 0 & -0.1 & -0.4 & 0 & 0 & 0 & 0 & 1 & -20/3 \\
\hline
6 & 0 & 0 & 0 & 0 & 0 & 0 & -0.2 & -0.5 & 0.3 & 0.1 & -0.2 & 0.4 & 1 & -20/3 \\
\hline
7 & 0 & 0 & 0 & 0 & 0 & 0 & 0.1 & -0.4 & 0.3 & 0 & -0.1 & 0 & 1 & -20/3 \\
\hline
8 & 0 & 0 & 0 & 0 & 0 & 0 & 0 & 0.2 & 0 & 0.4 & 0 & 0.1 & 1 & -20/3 \\
\hline
9 & 0 & 0 & 0 & 0 & 0 & 0 & 0.1 & -0.2 & 0.3 & 0 & -0.4 & 0.2 & 1.03 & -20/3 \\
\hline
10 & 0 & 0 & 0 & 0 & 0 & 0 & 0.1 & -0.4 & 0.3 & 0 & -0.1 & 0 & 1.01 & -20/3 \\
\hline
11 & 1 & 1 & 0 & 0 & 0 & 0 & 0 & 0 & 0 & 0 & 0 & 0 & 1 & -20/3 \\
\hline
12 & 0 & 0 & 1 & 1 & 0 & 0 & 0 & 0 & 0 & 0 & 0 & 0 & 1 & -20/3 \\
\hline
13 & 0 & 0 & 0 & 0 & 0 & 0 & 0.1 & -0.4 & 0.4 & 0 & -0.1 & 0 & 1 & -20/3 \\
\hline
14 & 0 & 0 & 0 & 0 & 0 & 0 & -0.2 & -0.5 & 0.3 & 0.1 & -0.2 & 0.3 & 1.01 & -20/3 \\
\hline
15 & 0 & 0 & 0 & 0 & 0 & 0 & -0.2 & -0.3 & 0 & 0 & 0 & 0 & 1 & -20/3 \\
\hline
16 & 0 & 0 & 0 & 0 & 0 & 0 & -0.2 & -0.5 & 0 & 0 & 0 & 0 & 1 & -20/3 \\
\hline
17 & 0 & 0 & 0 & 0 & 0 & 0 & -0.1 & -0.3 & 0 & 0 & 0 & 0 & 1 & -20/3 \\
\hline
18 & 0 & 0 & 0 & 0 & 0 & 0 & -0.1 & -0.2 & 0 & 0 & 0 & 0 & 1 & -20/3 \\
\hline
19 & 0 & 0 & 0 & 0 & 0 & 0 & -0.5 & 0.2 & 0 & 0 & 0 & 0 & 1 & -20/3 \\
\hline
20 & 0 & 0 & 0 & 0 & 0 & 0 & -0.2 & -0.4 & 0 & 0 & 0 & 0 & 1 & -20/3 \\
\hline
21 & 0 & 0 & 0 & 0 & 0 & 0 & -0.2 & -0.6 & 0 & 0 & 0 & 0 & 1 & -20/3 \\
\hline
22 & 0 & 0 & 0 & 0 & 0 & 0 & -0.3 & -0.5 & 0 & 0 & 0 & 0 & 1 & -20/3 \\
\hline
23 & 0 & 0 & 0 & 0 & 1 & 8 & 0 & 0 & 0 & 0 & 0 & 0 & 1 & -20/3 \\
\hline
24 & 1 & 8 & 0 & 0 & 0 & 0 & -0.2 & -0.5 & 0 & 0 & 0 & 0 & 1 & -7.75 \\
\hline
25 & 0.5 & 8 & 0 & 0 & 0 & 0 & -0.2 & -0.5 & 0 & 0 & 0 & 0 & 1 & -7.1 \\
\hline
\end{tabular}
\caption{Initial conditions (IC's) analyzed in this work (see Eq.\ (\ref{Bs0}) of the main text). IC's $\# 16$, $\# 21$, and $\# 22$ generate time evolutions which transiently violate the DEC at early times, when the system is far-from-equilibrium (these are the magenta, red, and salmon curves in Fig. \ref{fig:result1}, respectively). IC's $\# 23$ (originally proposed in \cite{wilkaodamassa}), $\# 24$, and $\# 25$ violate also the WEC (these are the blue, orange, and purple curves, respectively). All the IC's converge to the hydrodynamic regime at late times.}
\label{tabICs}
\end{table}

Regarding the definition of the subtracted fields in the bulk, they are motivated by the form of the UV expansions of the original bulk fields, namely
\begin{subequations}
\begin{align}
A(r,\tau) &= \frac{(r+\lambda(\tau))^2}{2}-\partial_\tau\lambda(\tau) + \frac{a_2(\tau)}{r^2} + \mathcal{O}(r^{-3}), \label{expA}\\
B(r,\tau) &= - \frac{2 \ln(\tau)}{3} - \frac{2}{3r\tau} + \frac{1+2\tau\lambda(\tau)}{3r^2 \tau^2} - \frac{2+6\tau\lambda(\tau)+6\tau^2\lambda^2(\tau)}{9r^3 \tau^3}\nonumber\\
& +\, \frac{6+24\tau\lambda(\tau)+36\tau^2\lambda^2(\tau)+24\tau^3\lambda^3(\tau)}{36r^4 \tau^4} - \frac{36\tau^4 a_2(\tau)+27\tau^5 \partial_\tau a_2(\tau)}{36r^4 \tau^4} + \mathcal{O}(r^{-5}),\label{expB}\\
\Sigma(r,\tau) &= \tau^{1/3}r + \frac{1+3\tau\lambda(\tau)}{3\tau^{2/3}} - \frac{1}{9r \tau^{5/3}} + \frac{5+9\tau\lambda(\tau)}{81r^2 \tau^{8/3}} - \frac{10+30\tau\lambda(\tau)+27\tau^2\lambda^2(\tau)}{243r^3 \tau^{11/3}} + \mathcal{O}(r^{-4}),\label{expS}\\
d_+\Sigma(r,\tau) &= \frac{\tau^{1/3}r^2}{2}+\frac{(1+3\tau\lambda(\tau))r}{3\tau^{2/3}} - \frac{1-2\tau\lambda(\tau)-3\tau^2\lambda^2(\tau)}{6\tau^{5/3}}+\frac{10}{81r \tau^{8/3}}\nonumber\\
& +\, \frac{-25-30\tau\lambda(\tau)+243\tau^4 a_2(\tau)}{243r^2 \tau^{11/3}} + \mathcal{O}(r^{-3}),\label{expdS}\\
d_+B(r,\tau) &= -\frac{1}{3\tau}+\frac{1}{3r\tau^2}-\frac{1+\tau\lambda(\tau)}{3r^2 \tau^3} + \frac{2\!+\!4\tau\lambda(\tau)\!+\!2\tau^2\lambda^2(\tau)\! +\!12\tau^4 a_2(\tau)\!+\!9\tau^5 \partial_\tau a_2(\tau)}{6 r^3 \tau^4}\! +\! \mathcal{O}(r^{-4}).\label{expdB}
\end{align}
\end{subequations}

We define the subtracted fields as follows, $u^p X_s(u,\tau)\equiv X(u,\tau) - X_{\textrm{UV}}(u,\tau)$, where $p$ is some integer and $X_{\textrm{UV}}$ is some UV truncation of $X$. Using this reasoning and looking at Eqs. \eqref{expA} --- \eqref{expdB}, we set here,
\begin{subequations}
\begin{align}
u^2 A_s(u,\tau) &\equiv A(u,\tau) - \frac{1}{2}\left(\frac{1}{u}+\lambda(\tau)\right)^2+\partial_\tau\lambda(\tau),\label{As}\\
u^4 B_s(u,\tau) &\equiv B(u,\tau) + \frac{2\ln(\tau)}{3} + \frac{2u}{3\tau} - \frac{(1+2\tau\lambda(\tau))u^2}{3\tau^2} + \frac{(2+6\tau\lambda(\tau)+6\tau^2\lambda^2(\tau))u^3}{9\tau^3},\label{Bs}\\
u^3 \Sigma_s(u,\tau) &\equiv \Sigma(u,\tau) -\frac{\tau^{1/3}}{u}-\frac{1+3\tau\lambda(\tau)}{3\tau^{2/3}}+\frac{u}{9\tau^{5/3}}  - \frac{(5+9\tau\lambda(\tau))u^2}{81\tau^{8/3}},\label{Ss}\\
u^2 (d_+\Sigma)_s(u,\tau) &\equiv d_+\Sigma(u,\tau)-\frac{\tau^{1/3}}{2u^2}-\frac{1+3\tau\lambda(\tau)}{3u\tau^{2/3}} + \frac{1-2\tau\lambda(\tau)-3\tau^2\lambda^2(\tau)}{6\tau^{5/3}} - \frac{10u}{81\tau^{8/3}},\label{dSs}\\
u^3 (d_+B)_s(u,\tau) &\equiv d_+B(u,\tau)+\frac{1}{3\tau}-\frac{u}{3\tau^2}+\frac{(1+\tau\lambda(\tau))u^2}{3\tau^3}.\label{dBs}
\end{align}
\end{subequations}
Then, the boundary values of the subtracted fields are simply given by radial constants,
\begin{subequations}
\begin{align}
A_s(u=0,\tau) &= a_2(\tau),\label{AAs}\\
B_s(u=0,\tau) &= -a_2(\tau)-\frac{3\tau\partial_\tau a_2(\tau)}{4}+\frac{1}{6\tau^4} + \frac{2\lambda(\tau)}{3\tau^3}+\frac{\lambda^2(\tau)}{\tau^2}+\frac{2\lambda^3(\tau)}{3\tau},\label{ABs}\\
\Sigma_s(u=0,\tau) &= -\frac{10+30\tau\lambda(\tau)+27\tau^2\lambda^2(\tau)}{243\tau^{11/3}},\label{ASs}\\
(d_+\Sigma)_s(u=0,\tau) &= \tau^{1/3} a_2(\tau)-\frac{25+30\tau\lambda(\tau)}{243\tau^{11/3}},\label{AdSs}\\
(d_+B)_s(u=0,\tau) &= 2a_2(\tau)+\frac{3\tau\partial_\tau a_2(\tau)}{2}+\frac{1}{3\tau^4} + \frac{2\lambda(\tau)}{3\tau^3} + \frac{\lambda^2(\tau)}{3\tau^2}.\label{AdBs}
\end{align}
\end{subequations}
In order to solve the equations of motions for the subtracted fields, one must rewrite Eqs. \eqref{pde1} --- \eqref{pde4} in terms of the new radial direction $u=1/r$ and also use Eqs. \eqref{As} --- \eqref{dBs} to express the original fields in terms of the subtracted ones.

We discretize the radial and time directions. The discretization of the radial domain of integration of the PDE's is implemented here by using the pseudospectral or collocation method \cite{boyd01}, with the discrete radial grid points described by the Chebyshev-Gauss-Lobatto grid,
\begin{equation}
u_k = \frac{u_{\textrm{IR}}}{2}\left[1+\cos\left(\frac{k\pi}{N-1}\right)\right], \ \ \ k=0,\dots , N-1,
\label{CBLgrid}
\end{equation}
where $N$ is the number of collocation points and $u_{\textrm{IR}}$ is the fixed infrared radial position in the interior of the bulk, from which one radially integrates the equations of motion up to the boundary at $u=0$. We used in this work $u_{\textrm{IR}}=1$ and $N = 33$.

The initial value of the dynamical UV coefficient $a_2(\tau)$ is one of the initial conditions which must be specified on the gravity side of the holographic duality (the other initial conditions being the initial profile of the bulk metric anisotropy, $B_s(u,\tau_0)$, and the initial value of the function $\lambda(\tau)$). In order to evolve $a_2(\tau)$ in time, one needs to determine the value of $\partial_\tau a_2(\tau)$. Knowing the values of $a_2(\tau)$ and $B_s(u=0,\tau)$, which on the initial time slice $\tau_0$ are freely chosen, one can determine the value of $\partial_\tau a_2(\tau)$ on the initial time $\tau_0$ using Eq. \eqref{ABs}.

In order to evolve the initial data $\{B_s(u,\tau),a_2(\tau);\lambda(\tau)\}$ in time, we also need to obtain $\partial_\tau B_s(u,\tau)$. This can be done by using Eq.\ \eqref{dBs} to relate the numerical field $(d_+B)_s(u,\tau)$ with $d_+B(u,\tau)=\partial_\tau B(u,\tau)-u^2 A(u,\tau) \partial_u B(u,\tau)$, and then expressing in this relation $A(u,\tau)$ and $B(u,\tau)$ in terms of the corresponding subtracted fields as given by Eqs.\ \eqref{As} and \eqref{Bs}. The resulting equation is solved for $\partial_\tau B_s(u,\tau)$ giving,
\begin{align}
\partial_\tau B_s(u,\tau) &= \frac{(d_+B)_s(u,\tau)}{u} -\frac{2}{3\tau^4 u} - \frac{2A_s(u,\tau)}{3\tau} + \frac{2uA_s(u,\tau)}{3\tau^2} - \frac{2u^2A_s(u,\tau)}{3\tau^3} + 4u^3A_s(u,\tau)B_s(u,\tau) + \frac{2B_s(u,\tau)}{u} \nonumber\\
& +\, \frac{B_s'(u,\tau)}{2} + u^4A_s(u,\tau)B_s'(u,\tau) + \left(4B_s(u,\tau)-\frac{2}{u\tau^3}+\frac{4u A_s(u,\tau)}{3\tau} - \frac{2u^2 A_s(u,\tau)}{\tau^2}+uB_s'(u,\tau)\right)\lambda(\tau)\nonumber\\
& +\! \left(\! - \frac{1}{3\tau^3} \!-\! \frac{7}{3\tau^2 u} \!+\! 2uB_s(u,\tau) \!-\! \frac{2u^2A_s(u,\tau)}{\tau} + \frac{u^2B_s'(u,\tau)}{2} \right)\lambda^2(\tau) - \left( \frac{1}{\tau^2} + \frac{4}{3\tau u} \right)\lambda^3(\tau) \nonumber\\
& -\, \frac{\lambda^4(\tau)}{\tau} + \left( \frac{2}{3\tau^3} - 4uB_s(u\tau) - u^2B_s'(u,\tau) + \frac{2\lambda(\tau)}{\tau^2} + \frac{2\lambda^2(\tau)}{\tau} \right)\partial_\tau\lambda(\tau),
\label{dtBs}
\end{align}
where $B_s'(u,\tau)\equiv \partial_u B_s(u,\tau)$ can be obtained at any constant time slice by simply applying the pseudospectral finite differentiation matrix \cite{Critelli:2017euk,boyd01} to the numerical solution $B_s(u,\tau)$.

Finally, we need to determine $\partial_\tau \lambda(\tau)$. We use the residual diffeomorphism invariance to set $\partial_\tau r_{\textrm{AH}}(\tau)=0$, which together with the requirement that \eqref{AppHor} holds at all times, imply that $\partial_\tau d_+\Sigma(r_{\textrm{AH}},\tau)=0$ and, thus, $d_+(d_+\Sigma)(r_{\textrm{AH}},\tau)=A\partial_r d_+\Sigma(r_{\textrm{AH}},\tau)$. Using this condition into the constraint equation \eqref{pde5}, and then combining the obtained result with the other components of Einstein's equations, one can show that the aforementioned requirements are realized by the following condition (already written in the radial coordinate $u=1/r$),
\begin{equation}
A(u_{\textrm{AH}},\tau) = 2\,\Sigma(u_{\textrm{AH}},\tau)\, \frac{3(d_+B(u_{\textrm{AH}},\tau))^2\Sigma(u_{\textrm{AH}},\tau)  +\, 6u_{\textrm{AH}}^2A'(u_{\textrm{AH}},\tau)d_+\Sigma(u_{\textrm{AH}},\tau)}{24\left[-u_{\textrm{AH}}^2\Sigma '(u_{\textrm{AH}},\tau)d_+\Sigma(u_{\textrm{AH}},\tau) - \Sigma^2(u_{\textrm{AH}},\tau)\right]} = -\frac{(d_+B(u_{\textrm{AH}},\tau))^2}{4},
\label{Astar}
\end{equation}
where above $'\equiv\partial_u$ and we used in the last step Eq. \eqref{AppHor}, $d_+\Sigma(u_{\textrm{AH}},\tau)=0$. Then, one can use Eq. \eqref{As} to obtain,
\begin{align}
\partial_\tau\lambda(\tau) = u_{\textrm{AH}}^2 A_s(u_{\textrm{AH}},\tau)+\frac{1}{2u_{\textrm{AH}}^2}+\frac{\lambda(\tau)}{u_{\textrm{AH}}}+\frac{\lambda^2(\tau)}{2} -A(u_{\textrm{AH}},\tau).
\label{dlambda}
\end{align}

Therefore, once a value for $\lambda(\tau)$ is chosen on the initial time slice $\tau_0$, $\lambda(\tau)$ can be evolved to the next time slice using Eq. \eqref{dlambda} such as to keep the radial position of the apparent horizon fixed during the time evolution. We set here the initial condition $\lambda(\tau_0)=0$ and solve Eq. \eqref{AppHor} using Eq. \eqref{dSs} and the Newton-Raphson algorithm. We find $u_{\textrm{AH}}$ off the collocation points with a good precision, given by the tolerance (or the number of iterations) of the method.

For the time evolution of the gravitational system we employed here the fourth-order Adams-Bashforth integration method with a time step size of $\Delta\tau = 12\times 10^{-5}$.

%%%%%%%%%%%%%%%%%%%%%%%%%%%%%%%%%%
\bibliographystyle{apsrev4-2}
\bibliography{Bibliography} % name of the bibtex file (in the same directory as the main tex file)

\end{document}